\documentclass[english]{article}
\usepackage[T1]{fontenc}
\usepackage[latin9]{inputenc}
\usepackage{geometry}
\usepackage{verbatim}
\usepackage{graphicx}
\usepackage{amssymb}
\usepackage{esint}

\makeatletter
\date{}

\makeatother

\usepackage{babel}

\begin{document}

\title{\textbf{{}Mathisson's helical motions demystified\textbf{}}}

\author{L. Filipe O. Costa%
\thanks{filipezola@fc.up.pt%
}, José Natário%
\thanks{jnatar@math.ist.utl.pt%
}\textbf{, }Miguel Zilhão%
\thanks{mzilhao@fc.up.pt%
} \\
 \\
 {\em {\small $^{*\ddagger}$Centro de F\'{\i}sica do Porto, Universidade
do Porto} }\\
{\small{} {\em Rua do Campo Alegre, 687, 4169-007 Porto, Portugal}}\\
{\small{} {\em $^{\dagger}$Departamento de Matemática, Instituto
Superior Técnico, 1049-001, Lisboa, Portugal}}}

\maketitle
\begin{abstract}
The motion of spinning test particles in general relativity is described
by Mathisson-Papapetrou-Dixon equations, which are undetermined up
to a spin supplementary condition, the latter being today still an
open question. The Mathisson-Pirani (MP) condition is known to lead
to rather mysterious helical motions which have been deemed unphysical,
and for this reason discarded. We show that these assessments are
unfounded and originate from a subtle (but crucial) misconception.
We discuss the kinematical explanation of the helical motions, and
dynamically interpret them through the concept of hidden momentum,
which has an electromagnetic analogue. We also show that, contrary
to previous claims, the frequency of the helical motions coincides
exactly with the zitterbewegung frequency of the Dirac equation for
the electron. \textbf{}
\end{abstract}

\section{Introduction. Mathisson's helical motions.}

(Note: for an explanation of the notation herein, see \cite{Helical}).
In a multipole expansion, a body is represented by a set of moments
of its energy-momentum tensor $T^{\alpha\beta}$, taken about a reference
worldline $z^{\alpha}(\tau)$. Spinning pole-dipole particles correspond
to truncating the expansion at dipole order; the equations of motion
resulting from $T_{\ \ ;\beta}^{\alpha\beta}=0$ involve only two
moments of $T^{\alpha\beta}$: the momentum $P^{\alpha}$, and the
angular momentum $S^{\alpha\beta}$ (see definitions in \cite{Helical});
and read for a free particle in flat spacetime:\begin{equation}
\frac{DP^{\alpha}}{d\tau}=0\quad{\rm (a)};\quad\frac{DS^{\alpha\beta}}{d\tau}=2P^{[\alpha}U^{\beta]}\quad{\rm (b)};\quad P^{\alpha}=mU^{\alpha}-\frac{DS^{\alpha\beta}}{d\tau}U_{\beta}\quad{\rm (c)},\label{eq:Eqs Motion Flat}\end{equation}
 (\ref{eq:Eqs Motion Flat}c) following from (\ref{eq:Eqs Motion Flat}b);
$U^{\alpha}=dz^{\alpha}/d\tau$ and $m\equiv-P^{\alpha}U_{\alpha}$
is the proper mass. There are three more unknowns than equations;
to form a determined system, these equations require thus a supplementary
condition, which amounts to specifying $z^{\alpha}(\tau)$. Mathisson's
helical solutions~\cite{Mathisson Zitterbewegung} arise when one
uses the condition $S^{\alpha\beta}U_{\alpha}=0$. In this case $S^{\alpha\beta}=\epsilon^{\alpha\beta\mu\nu}S_{\mu}U_{\nu}$;
(\ref{eq:Eqs Motion Flat}c) becomes $P^{\alpha}=mU^{\alpha}+S^{\alpha\beta}a_{\beta}$,
where $a^{\alpha}=DU^{\alpha}/d\tau$; and $DS^{\alpha}/d\tau=0$.
The general solution of (\ref{eq:Eqs Motion Flat}) under this condition
describes the famous helical motions, which, \emph{in the} $P^{i}=0$
\emph{frame}, correspond to \textit{clockwise} (i.e.~\emph{opposite}
to the spin direction) circular motions with radius $R=v\gamma^{2}S/m$
and speed $v$ on the $xy$ plane; taking their center as the spatial
origin of the frame, they read: \begin{equation}
z^{\alpha}(\tau)=\left(\gamma\tau,-R\cos\left(\frac{v\gamma}{R}\tau\right),R\sin\left(\frac{v\gamma}{R}\tau\right),0\right)\label{eq:PositionMathisson}\end{equation}
 These motions were interpreted \cite{Mathisson Zitterbewegung} (for
the case of the electron) as the classical counterpart of Dirac's
equation `zitterbewegung'. However, the fact that $\gamma$ can be
arbitrarily large has led some authors~(see e.g. \cite{WeyssenhoffRaabe,Dixon1965}),
to believe%
{}{} that according to (\ref{eq:PositionMathisson}) a given free body
might move along circular trajectories with any radius; for this reason
these solutions have been deemed \emph{unphysical}. The same arguments
were used to imply that the the frequency $\omega=m/\gamma^{2}S$,
for an electron, only coincides with Dirac's zitterbewegung frequency
$\omega=2M_{e}/\hbar$ in the limit $\gamma\rightarrow1$. Both these
assessments are misconceptions as we will see next.

\section{Center of mass. Significance of the spin condition. Kinematical origin
of the helical motions. \label{sub:Center-of-mass}}

In order for (\ref{eq:Eqs Motion Flat}) to be equations of motion
for the body, $z^{\alpha}(\tau)$ must be taken as its representative
point --- its center of mass (CM); however, in relativity, the CM
of a spinning body is an observer dependent point. This is illustrated
in Fig. 1 of \cite{Helical}. A spin condition of the type $S^{\alpha\beta}u_{\beta}=0$
(for some unit time-like vector $u^{\alpha}$) amounts to choosing
$z^{\alpha}(\tau)$ as the center of mass $x_{{\rm CM}}^{\alpha}(u)$
measured by the observer $\mathcal{O}(u)$ of 4-velocity $u^{\alpha}$,
see \cite{Helical} for details. %
{}{}The Mathisson-Pirani condition $S^{\alpha\beta}U_{\alpha}=0$ amounts
to choosing for $z^{\alpha}$ the center of mass $x_{{\rm CM}}^{\alpha}(U)$
as measured in its own rest frame, i.e., the frame $U^{i}=0$. Such
CM is dubbed a {}``proper center of mass''. It turns out that, contrary
to what one might expect, such point is not unique. Let $x_{{\rm CM}}^{\alpha}(P)$
be the CM measured in the $P^{i}=0$ frame; for a free particle in
flat spacetime it is one of the proper CM's, corresponding to $R=0$
in Eq. (\ref{eq:PositionMathisson}). This solution corresponds to
uniform straightline motion. The center of mass $x_{{\rm CM}}^{\alpha}(\bar{u})$
measured by an observer $\mathcal{O}(\bar{u})$ moving with 3-velocity
$\vec{v}$ in the $P^{i}=0$ frame %
{}{}is shifted by a vector $\Delta x^{i}$, Eq. (\ref{eq:ShiftVel}a),
relative to $x_{{\rm CM}}^{i}(P)$. Hence the set of all possible
CM's measured by all observers $\mathcal{O}(\bar{u})$ fills a disk
of radius $R_{max}=S_{*}/M$ centered at $x_{{\rm CM}}^{\alpha}(P)$.
\begin{equation}
\Delta x^{i}=\ \frac{(\vec{S}_{\star}\times\vec{v})^{i}}{M}\;(a);\quad\frac{D\Delta x^{\alpha}}{d\tau_{P}}=-\frac{S_{\star}^{\alpha\beta}}{M}\frac{Dv_{\beta}}{d\tau_{P}}\ \;(b)\quad\Leftrightarrow\ \frac{d\vec{\Delta x}}{dt}=\frac{1}{M}\vec{S}_{\star}\times\frac{d\vec{v}}{dt}\ \;(c).\label{eq:ShiftVel}\end{equation}
 Note: $M\equiv\sqrt{-P^{\alpha}P_{\alpha}}$; $S_{\star}^{\alpha\beta}$
is the angular momentum taken about $x_{{\rm CM}}^{\alpha}(P)$. If
$\mathcal{O}(\bar{u})$ is inertial, $x_{{\rm CM}}^{\alpha}(\bar{u})$
is a point at rest relative to $x_{{\rm CM}}^{\alpha}(P)$, c.f. Eqs.
(\ref{eq:ShiftVel}b)-(\ref{eq:ShiftVel}c); thus \emph{not} at rest
relative to $\mathcal{O}(\bar{u})$, i.e., it is \emph{not} a proper
CM. But if $\vec{v}$ is not constant, then $x_{{\rm CM}}^{\alpha}(\bar{u})$
acquires a non-trivial velocity $\vec{v}_{{\rm CM}}=d\vec{\Delta x}/dt$
(as measured in the $P^{i}=0$ frame). If $\mathcal{O}(\bar{u})$
itself moves with $\vec{v}=\vec{v}_{{\rm CM}}$, i.e. if $\vec{v}$
is a solution of Eq. (\ref{eq:ShiftVel}c), then it is a proper CM
(i.e., it is a CM at rest relative to the frame where it is computed).
The solutions (in the $P^{i}=0$ frame) are circular motions in the
plane orthogonal to $\vec{S}_{\star}$, with radius $R=\Delta x=|\vec{v}\times\vec{S}_{\star}|/M$,
and constant (independent of $R$) angular velocity $\omega=-M/S_{\star}$
in \emph{opposite sense} to the rotation of the body. These are precisely
the solutions (\ref{eq:PositionMathisson}), and this is origin of
the helical motions \cite{MollerAIP}%
{}{}. Hence their radius is not arbitrarily large; they are contained
within the disk of CM's, of radius $R_{max}=S_{\star}/M$; which is
actually the minimum size a particle can have without violating the
dominant energy condition (i.e., without possessing matter/energy
flowing faster than light). The latter implies $\rho>|\vec{J}|$,
where $\rho\equiv T^{00}$ and $J^{i}\equiv T^{0i}$; let $b$ be
the largest dimension of the body. Using the definition of $S_{\star}^{\alpha\beta}$
in \cite{Helical}, we may write, in the $P^{i}=0$ frame: \[
S_{\star}=\left|\int\vec{r}\times\vec{J}d^{3}x\right|\le\int r|\vec{J}|d^{3}x<\int\rho rd^{3}x\le Mb\ \Leftrightarrow\ b>\frac{S_{\star}}{M}=R_{max}\]
 Thus the disk of CM's, within which all the helical motions are contained,
is always smaller than the body.

\emph{The misconception in the literature.} --- Different representations
of the same extended body must yield the same moments ($P^{\alpha}$
and $S^{\alpha\beta}$) with respect to the \emph{same observer} and
the \emph{same reference worldline}. As shown in \cite{Helical},
it is the quantities $S_{\star}=\gamma S$ and $M=m/\gamma$, not
$m$ and $S$ (which depend, via $U^{\alpha}$ and $z^{\alpha}$,
respectively, on the particular helix chosen), that we must fix in
order to ensure that we are dealing with the same particle. Thus,
$R=v\gamma^{2}S/m=vS_{\star}/M\le R_{max}$, for all the helical representations
corresponding to a given particle. Moreover, the frequency $\omega=m/\gamma^{2}S=M/S_{\star}$
is the same for all helices corresponding to the same particle, and
coincides exactly (even in the relativistic limit) with Dirac's zitterbewegung
frequency, identifying $S_{\star}=\hbar/2$ and $M=M_{e}$.

\section{Dynamical Interpretation of the Helical Motions}

\begin{figure}
\includegraphics[width=1\textwidth]{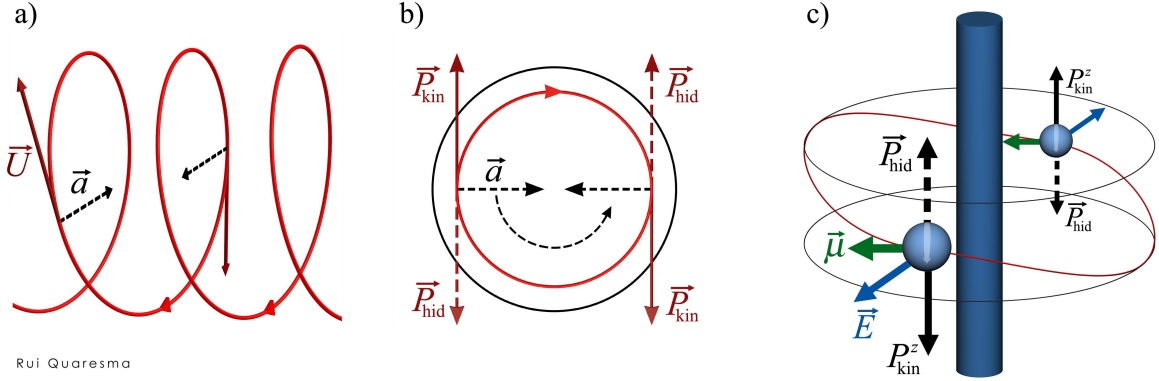}

\caption{\label{fig:HiddenHelical}Hidden momentum provides dynamical interpretation
for the helical motions: the acceleration results from an interchange
between kinetic $P_{kin}^{\alpha}=mU^{\alpha}$ and hidden {}``inertial''
momentum $P_{hid}^{\alpha}=S^{\alpha\beta}a_{\beta}$, which occurs
in a way that their variations cancel out at every instant, keeping
$P^{\alpha}$ constant. This is made manifest in b) panel, representing
the $\vec{P}=0$ frame, wherein $\vec{P}_{hid}=\vec{a}\times_{U}\!\vec{S}=-m\vec{U}=-\vec{P}_{kin}$.
Panel c) represents an electromagnetic analogue \cite{Wald et al 2010}:
a (negatively) charged test particle possessing magnetic dipole moment
$\vec{\mu}=(\mu^{x},\mu^{y},0)$, orbiting a cylindrical (positively)
charged body. The cylinder is along the $z$ axis, and $\vec{E}$
is the electric field it produces (measured in the particle's CM frame).
The $z$ component of the force vanishes for this setup; hence $P^{z}=0=constant$.
But the particle possesses a hidden momentum \cite{Wald et al 2010,Gyros}
$\vec{P}_{hid}=\vec{\mu}\times_{U}\!\vec{E}$; as it orbits the line
charge, $\vec{P}_{hid}$ oscillates between positive and negative
values along the $z$-axis, implying the particle to bob up and down
in order to keep the total momentum along $z$ constant: $P^{z}=P_{kin}^{z}+P_{hid}^{z}=0$.
(Note however the important distinction: $\vec{a}\times_{U}\!\vec{S}$,
but \emph{not} $\vec{\mu}\times_{U}\!\vec{E}$, is pure gauge).}

\end{figure}

We see from Eq. (\ref{eq:ShiftVel}b) that the CM $x_{{\rm CM}}^{\alpha}(u)$
is not at rest in the $\vec{P}=0$ frame when the 4-velocity $u^{\alpha}$
of the observer measuring it changes; conversely, $\vec{P}$ will
not be zero in the CM frame (where, by definition, the particle is
at rest); thus $P^{\alpha}$ is not parallel to $U^{\alpha}$, and
the particle is said to possess hidden momentum \cite{Wald et al 2010}.
This is a key concept for the understanding of the dynamics of the
helical solutions; namely how the CM of a spinning particle can accelerate
in the absence of any force without violating the conservation laws.
Consider a generic spin condition $S^{\alpha\beta}u_{\beta}=0$; contracting
(\ref{eq:Eqs Motion Flat}b) with $u_{\beta}$, leads to \begin{equation}
\ S^{\alpha\beta}\frac{Du_{\beta}}{d\tau}=\gamma(u,U)P^{\alpha}-m(u)U^{\alpha};\label{eq:HiddenInertial}\end{equation}
 where $\gamma(u,U)\equiv-U^{\beta}u_{\beta}$ and $m(u)\equiv-P^{\beta}u_{\beta}$.
We split the momentum $P^{\alpha}$ in two parts: {}``kinetic momentum''
$P_{{\rm kin}}^{\alpha}=mU^{\alpha}$, which is the projection of
$P^{\alpha}$ along $U^{\alpha}$; and the projection orthogonal to
$U^{\alpha}$, $P_{{\rm hid}}^{\alpha}\equiv(h^{U})_{\ \beta}^{\alpha}P^{\beta}$,
which is the hidden momentum. Hence, if $Du_{\beta}/d\tau=0$, that
is, if we take as $z^{\alpha}(\tau)$ the CM measured by an observer
$\mathcal{O}(u)$ \emph{such that} $u^{\alpha}$ \emph{is parallel
transported along it} (e.g., an inertial observer in flat spacetime),
then $P^{\alpha}\parallel U^{\alpha}$, and $P_{{\rm hid}}^{\alpha}=0$.
Otherwise, $P_{{\rm hid}}^{\alpha}\ne0$ in general. This is reciprocal
to Eq. (\ref{eq:ShiftVel}b), one can obtain one effect from the other,
see \cite{Helical}. Notice the important message encoded herein:
in relativity, the motion of a spinning particle is not determined
by the force laws given the initial position and velocity; one needs
also to determine the field of vectors $u^{\alpha}$ relative to which
the CM is computed; the variation of $u^{\alpha}$ along $z^{\alpha}(\tau)$
is enough to possibly cause the CM to accelerate, even in the absence
of any force; in this case the variation of $P_{{\rm kin}}^{\alpha}$
is compensated by an opposite variation of $P_{{\rm hid}}^{\alpha}$,
keeping $P^{\alpha}$ constant. If $u^{\alpha}$ varies in a way such
that the signal in Eq. (\ref{eq:ShiftVel}b) oscillates, we may have
a bobbing; or if it is such that $\mathcal{O}(u)$ sees its CM to
be at rest ($u^{\alpha}=U^{\alpha}$, i.e, its 3-velocity $\vec{v}$,
in the frame $P^{i}=0$, is a solution of $\vec{v}=\vec{v}_{{\rm CM}}$,
Eq. (\ref{eq:ShiftVel}c)), so that the condition $S^{\alpha\beta}u_{\beta}=S^{\alpha\beta}U_{\beta}=0$
is obeyed, then we have a helical solution. In this case $P_{hid}^{\alpha}=S^{\alpha\beta}a_{\beta}=\epsilon_{\ \beta\gamma\delta}^{\alpha}a^{\beta}S^{\gamma}U^{\delta}$,
which in vector notation reads $\vec{P}_{hid}=-\vec{S}\times_{U}\!\vec{a}=\vec{S}\times_{U}\!\vec{G}$
where $\vec{G}$ is the {}``gravitoelectric field'' as measured
in the CM frame \cite{Gyros}. This is formally analogous to the hidden
momentum $P_{{\rm hid}}^{\alpha}=\epsilon_{\ \beta\gamma\delta}^{\alpha}E^{\gamma}\mu^{\beta}U^{\delta}$
of electromagnetic systems; in vector notation $\vec{P}_{{\rm hid}}=\vec{\mu}\times_{U}\!\vec{E}$,
see e.g. \cite{Gyros,Wald et al 2010}. The dynamics of the helical
representations may be cast as analogous to the bobbing \cite{Wald et al 2010}
of a magnetic dipole orbiting a cylindrical charge, as explained in
Fig. \ref{fig:HiddenHelical}.%

Concluding: there is nothing wrong with Mathisson-Pirani condition,
it is as valid as any other of the (infinite) possible choices; and
in some applications the most suitable one, see \cite{Gyros}. It
is degenerate, and the helical solutions it allows for a free particle,
in addition to the expected uniform straightline motion, are \emph{alternative}
and \emph{physically consistent} descriptions of the motion (only
more complicated): in the first case, we have $a^{\alpha}=P_{{\rm hid}}^{\alpha}=0$;
in the second case we have an helix, but also $P_{{\rm hid}}^{\alpha}\ne0$
(the latter being pure gauge and the motion effects induced by it
confined to the worldtube of CM's).

\end{document}